\documentclass[12pt,titlepage]{article}

\usepackage{amssymb}
\usepackage[dvips]{graphicx}
\usepackage{amsfonts}
\usepackage[portuguese]{babel}

\begin{document}

\title{\textbf{Espalhamento e estados ligados}\\
\textbf{\ em potenciais localizados}\\
{\small (\textbf{Scattering and bound states by localized potentials})}}
\author{A.S. de Castro\thanks{%
E-mail: castro@pq.cnpq.br} \\
\\
Departamento de F\'{\i}sica e Qu\'{\i}mica\\
Universidade Estadual Paulista\\
Guaratinguet\'{a} SP - Brasil}
\date{}
\maketitle

\begin{abstract}
Apresenta-se um formalismo simples que permite explorar o espalhamento qu%
\^{a}ntico e os poss\'{\i}veis estados ligados em um potencial sim\'{e}trico
localizado de forma arbitr\'{a}ria de um modo unificado. A barreira e o po%
\c{c}o quadrados sim\'{e}tricos s\~{a}o utilizados como ilustra\c{c}\~{a}o
do m\'{e}todo.\newline
\newline
\noindent \textbf{Palavras-chave:} espalhamento, estado ligado, potencial
localizado, coeficiente de transmiss\~{a}o.\newline
\newline
\newline

\noindent {\small {A simple formalism for exploring quantum scattering and
possible bound states in an arbitrary symmetric and localized potential in a
unified way is presented. The symmetric square barrier and well potentials
are used for illustrating the method. \newline
\newline
}}

{\small {\noindent \textbf{Keywords:} \ scattering, bound state, localized
potential, transmission coefficient.}}
\end{abstract}

\section{Introdu\c{c}\~{a}o}

Um exame detalhado do espalhamento qu\^{a}ntico em um potencial retangular
generalizado foi publicado recentemente nesta Revista por C\^{a}ndido
Ribeiro e colaboradores \cite{can}. Nesse estudo, ap\'{o}s uma proficiente
descri\c{c}\~{a}o das aplica\c{c}\~{o}es do espalhamento qu\^{a}ntico, desde
o decaimento alfa at\'{e} os \textit{quantum dots}, os autores exploram um
potencial retangular constitu\'{\i}do de tr\^{e}s patamares que reduz-se ao
po\c{c}o de potencial, \`{a} barreira de potencial e ao degrau duplo,
consoante o ajuste de dois par\^{a}metros do potencial generalizado. O
coeficiente de transmiss\~{a}o \'{e} calculado exatamente, e alguns casos
particulares, incluindo po\c{c}os e barreiras assim\'{e}tricos, s\~{a}o
estudados com certa min\'{u}cia.

O presente trabalho apresenta um formalismo simples que permite explorar os
estados de espalhamento, tanto quanto os poss\'{\i}veis estados ligados, em
um potencial sim\'{e}trico lo\-ca\-li\-za\-do de forma arbitr\'{a}ria. O m%
\'{e}todo permite abordar o problema de espalhamento e estados ligados de
uma forma unificada utilizando-se de um ferramental matem\'{a}tico acess%
\'{\i}vel aos estudantes de f\'{\i}sica j\'{a} nos cursos introdut\'{o}rios
de mec\^{a}nica qu\^{a}ntica. A barreira e o po\c{c}o quadrados sim\'{e}%
tricos, problemas analiticamente sol\'{u}veis que se fazem presentes nos
livros-texto de mec\^{a}nica qu\^{a}ntica, s\~{a}o utilizados como ilustra%
\c{c}\~{a}o do m\'{e}todo.

\section{Solu\c{c}\~{a}o para um potencial localizado}

A equa\c{c}\~{a}o de Schr\"{o}dinger unidimensional para uma part\'{\i}cula
de massa de repouso $m$ sujeita a um potencial $V(x,t)$ \'{e} dada por%
\begin{equation}
i\hbar \,\frac{\partial \Psi (x,t)}{\partial t}=-\frac{\hbar ^{2}}{2m}\,%
\frac{\partial ^{2}\Psi (x,t)}{\partial x^{2}}+V(x,t)\Psi (x,t)  \label{SCH}
\end{equation}%
onde $\hbar $ \'{e} a constante de Planck dividida por $2\pi $, e $\Psi
(x,t) $ \'{e} a fun\c{c}\~{a}o de onda. A equa\c{c}\~{a}o da continuidade
para a equa\c{c}\~{a}o de Schr\"{o}dinger%
\begin{equation}
\frac{\partial \rho }{\partial t}+\frac{\partial J}{\partial x}=0
\label{con}
\end{equation}%
\noindent \'{e} satisfeita com $\rho $ e $J$ definidos como%
\begin{equation}
\rho =|\Psi |^{2},\quad J=\frac{\hbar }{2im}\left( \Psi ^{\ast }\frac{%
\partial \Psi }{\partial x}-\frac{\partial \Psi ^{\ast }}{\partial x}\Psi
\right)  \label{rho}
\end{equation}%
A grandeza $\rho $ \'{e} interpretada como uma densidade de probabilidade e $%
J$ como uma corrente (ou fluxo) de probabilidade. Para um potencial
independente do tempo, equa\c{c}\~{a}o de Schr\"{o}dinger admite solu\c{c}%
\~{o}es da forma%
\begin{equation}
\Psi (x,t)=\psi (x)\,\,e^{-i\frac{E}{\hbar }t}  \label{it}
\end{equation}

\noindent onde $\psi $ obedece \`{a} equa\c{c}\~{a}o de Schr\"{o}dinger
independente do tempo

\begin{equation}
\left( -\frac{\hbar ^{2}}{2m}\,\frac{d^{2}}{dx^{2}}+V\right) \,\psi =E\,\psi
\label{3}
\end{equation}%
e a densidade e corrente correspondentes \`{a} solu\c{c}\~{a}o expressa por (%
\ref{it}) tornam-se%
\begin{equation}
\rho =\left\vert \psi \right\vert ^{2},\quad J=\frac{\hbar }{2im}\left( \psi
^{\ast }\frac{\partial \psi }{\partial x}-\frac{\partial \psi ^{\ast }}{%
\partial x}\psi \right)  \label{rho1}
\end{equation}%
Em virtude de $\rho $ e $J$ serem independentes do tempo, a solu\c{c}\~{a}o (%
\ref{it}) \'{e} dita descrever um estado estacion\'{a}rio. Tamb\'{e}m, a lei
de conserva\c{c}\~{a}o expressa por (\ref{con}) implica que o fluxo de
probabilidade \'{e} independente de $x$ para os estados estacion\'{a}rios.

Vamos agora considerar a equa\c{c}\~{a}o de Schr\"{o}dinger com um potencial
independente do tempo localizado. O potencial localizado, n\~{a}o-nulo
apenas numa regi\~{a}o finita do eixo $x$, \'{e} expresso como
\begin{equation}
V(x)=\mathcal{V}(x)\left[ \theta \left( x+L\right) -\theta \left( x-L\right) %
\right] =\left\{
\begin{array}{c}
0 \\
\\
\mathcal{V}(x)%
\end{array}%
\begin{array}{c}
{\textrm{para }}|x|>L \\
\\
{\textrm{para }}|x|<L%
\end{array}%
\right.  \label{4}
\end{equation}%
onde $\theta (x)$ \'{e} a fun\c{c}\~{a}o de Heaviside:%
\begin{equation}
\theta (x)=\left\{
\begin{array}{c}
0 \\
\\
1%
\end{array}%
\begin{array}{c}
{\textrm{para }}x<0 \\
\\
{\textrm{para }}x>0%
\end{array}%
\right.  \label{heav}
\end{equation}

Para $x<-L$, a equa\c{c}\~{a}o de Schr\"{o}dinger apresenta a solu\c{c}\~{a}%
o geral%
\begin{equation}
\psi =a_{+}\,e^{+ikx}+a_{-}\,e^{-ikx}  \label{5}
\end{equation}%
onde o n\'{u}mero de onda $k$ \'{e} definido como
\begin{equation}
k=\sqrt{\frac{2m}{\hbar ^{2}}E}  \label{6}
\end{equation}%
Para $E>0$, a solu\c{c}\~{a}o expressa por (\ref{5}) reverte-se em uma soma
de autofun\c{c}\~{o}es do operador momento ($p_{\mathtt{op}}=-i\hbar
\partial /\partial x$). Tais autofun\c{c}\~{o}es descrevem ondas planas
pro\-pa\-gan\-do-se em ambos os sentidos do eixo $x$ com velocidade de grupo
(veja, e.g., Ref. \cite{gri})%
\begin{equation}
v_{g}=\frac{1}{\hbar }\,\frac{dE}{dk}  \label{7}
\end{equation}%
igual \`{a} velocidade cl\'{a}ssica da part\'{\i}cula. Por conseguitnte, $%
a_{+}\,e^{+ikx}$ descreve par\-t\'{\i}\-cu\-las incidentes ($v_{g}=\hbar
k/m>0$), enquanto $a_{-}\,e^{-ikx}$ descreve part\'{\i}culas
re\-fle\-ti\-das ($v_{g}=-\hbar k/m<0$). A corrente nesta regi\~{a}o do espa%
\c{c}o, correspondendo a $\psi $ dada por \ (\ref{5}), \'{e} expressa por
\begin{equation}
J=J_{\mathtt{inc}}-J_{\mathtt{ref}}  \label{7a}
\end{equation}%
onde%
\begin{equation}
J_{\mathtt{inc}}=\frac{\hbar k}{m}\,|a_{+}|^{2},\quad J_{\mathtt{ref}}=\frac{%
\hbar k}{m}\,|a_{-}|^{2}  \label{70a}
\end{equation}%
Observe que a rela\c{c}\~{a}o $J=\rho \,v_{g}$ mant\'{e}m-se tanto para a
onda incidente quanto para a onda refletida, pois%
\begin{equation}
\rho _{\pm }=\,|a_{\pm }|^{2}  \label{7b}
\end{equation}

Por outro lado, para $x>L$ as solu\c{c}\~{o}es s\~{a}o da forma%
\begin{equation}
\psi =b_{+}\,e^{+ikx}+b_{-}\,e^{-ikx}  \label{7c}
\end{equation}%
Para termos uma onda progressiva se afastando da regi\~{a}o do potencial
(pro\-pa\-gan\-do-se no sentido positivo do eixo $x$ com $v_{g}=\hbar k/m>0$%
) devemos impor $b_{-}=0$. A densidade e a corrente nesta regi\~{a}o do espa%
\c{c}o, correspondendo a $\psi $ dada por \ (\ref{7c}) com $b_{-}=0$, s\~{a}%
o expressas por%
\begin{equation}
\rho =\,|b_{+}|^{2},\quad J_{\mathtt{trans}}=\frac{\hbar k}{m}\,|b_{+}|^{2}
\label{70c}
\end{equation}

Para $|x|<L$ a solu\c{c}\~{a}o geral tem a forma%
\begin{equation}
\psi =c_{P}\,u(x)+c_{I}\,v(x)  \label{8}
\end{equation}%
onde $u$ e $v$ s\~{a}o solu\c{c}\~{o}es linearmente independentes da equa%
\c{c}\~{a}o de Schr\"{o}dinger, e $c_{P}$ e $c_{I}$ s\~{a}o constantes arbitr%
\'{a}rias. Doravante, por motivos de simplicidade, vamos considerar um
potencial par, i.e. $\mathcal{V}(-x)=+\mathcal{V}(x)$, de modo que podemos
considerar solu\c{c}\~{o}es com paridade definida\footnote{%
Se $\phi (x)$ satisfaz \`a equa\c{c}\~{a}o de Schr\"{o}dinger independente
do tempo para um dado $E$, assim acontece com $\phi (-x)$, e portanto tamb%
\'{e}m satisfazem as combina\c{c}\~{o}es lineares $\phi (x)\pm \phi (-x)$.}.
Seja $u$ a solu\c{c}\~{a}o par e $v$ a solu\c{c}\~{a}o \'{\i}mpar:%
\begin{equation}
u(-x)=u(x)\quad \textrm{e}\quad v(-x)=-v(x)  \label{parimpar}
\end{equation}%
Mais ainda, sem perda de generalidade, podemos considerar que $u$ e $v$ s%
\~{a}o fun\c{c}\~{o}es reais\footnote{%
Se $\phi $ satisfaz \`a equa\c{c}\~{a}o de Schr\"{o}dinger independente do
tempo para um dado $E$, assim acontece com $\phi ^{\ast }$, e portanto tamb%
\'{e}m satisfazem as combina\c{c}\~{o}es lineares $\phi \pm \phi ^{\ast }$.}%
. Neste caso, o leitor pode facilmente verificar que%
\begin{equation}
J\left( |x|<L\right) =\frac{\hbar W(x)}{m}\,\textrm{Im}\left( c_{P}^{\ast
}\,c_{I}\right)  \label{jota}
\end{equation}%
onde $W$ \'{e} o wronskiano das solu\c{c}\~{o}es $u$ e $v$, i.e. $%
W=u\,v^{\prime }-u^{\prime }\,v$, onde a plica $^{\prime }$ (tamb\'{e}m
conhecida como linha, irreconhec\'{\i}vel como \textit{primo} na L\'{\i}ngua
Portuguesa) significa a derivada em rela\c{c}\~{a}o a $x$. Sucede que o
wronskiano para duas solu\c{c}\~{o}es linearmente independentes de uma equa%
\c{c}\~{a}o diferencial de segunda ordem \'{e} diferente de zero, e para o
caso a equa\c{c}\~{a}o de Schr\"{o}dinger, como o leitor pode verificar,
\'{e} independente de $x$. Assim, podemos at\'{e} mesmo escrever $%
W=u(0)v^{\prime }(0)$.

Come\c{c}aremos agora o c\'{a}lculo de grandezas de suma import\^{a}ncia na
des\-cri\-\c{c}\~{a}o do es\-pa\-lha\-men\-to, viz. os coeficientes de reflex%
\~{a}o e transmiss\~{a}o. Assim, $k$, definido em (\ref{6}), \'{e} uma
quantidade real. N\~{a}o obstante poss\'{\i}veis descontinuidades do
potencial em $x_{0}=\pm L$, a autofun\c{c}\~{a}o e sua derivada primeira s%
\~{a}o fun\c{c}\~{o}es cont\'{\i}nuas. Esta conclus\~{a}o, v\'{a}lida para
potenciais com descontinuidades finitas, pode ser obtida pela integra\c{c}%
\~{a}o da Eq. (\ref{3}) entre $x_{0}-\varepsilon $ e $x_{0}+\varepsilon $ no
limite $\varepsilon \rightarrow 0$. Pode-se verificar, pelo mesmo
pro\-ce\-di\-men\-to, que apenas as autofun\c{c}\~{o}es s\~{a}o cont\'{\i}%
nuas quando as descontinuidades dos potenciais s\~{a}o infinitas.

A demanda por continuidade de $\psi $ e $d\psi /dx$ fixa todas as amplitudes
em termos da amplitude da onda incidente $a_{+}$. A continuidade em $x=-L$
\'{e} expressa como%
\begin{eqnarray}
a_{+}\,e^{-ikL}+a_{-}\,e^{+ikL} &=&c_{P}\,u_{L}-c_{I}v\,_{L}  \nonumber \\
&&  \label{C1} \\
ik\left( a\,_{+}\,e^{-ikL}-a_{-}\,e^{+ikL}\right) &=&-c_{P}\,u_{L}^{\prime
}+c_{I}\,v_{L}^{\prime }  \nonumber
\end{eqnarray}%
e em $x=+L$ como%
\begin{eqnarray}
b_{+}\,e^{+ikL} &=&c_{P}\,u_{L}+c_{I}\,v_{L}  \nonumber \\
&&  \label{C2} \\
ikb_{+}\,e^{+ikL} &=&c_{P}\,\,u_{L}^{\prime }+c_{I\,}v_{L}^{\prime }
\nonumber
\end{eqnarray}%
onde o subscrito $L$ em $u$ e $v$ significa avalia\c{c}\~{a}o em $x=L$. De (%
\ref{C1}) e (\ref{C2}), temos
\begin{equation}
\frac{c_{P}}{a_{+}}=-\,\frac{ike^{-ikL}}{u_{L}^{\prime }-iku_{L}}
\label{13c}
\end{equation}

\begin{equation}
\frac{c_{I}}{a_{+}}=+\,\frac{ike^{-ikL}}{v_{L}^{\prime }-ikv_{L}}
\label{13d}
\end{equation}

\begin{equation}
\frac{a_{-}}{a_{+}}=-\,\frac{e^{-2ikL}\left( k^{2}u_{L}\,v_{L}+u_{L}^{\prime
}\,v_{L}^{\prime }\right) }{\left( u_{L}^{\prime }-iku_{L}\right) \left(
v_{L}^{\prime }-ikv_{L}\right) }  \label{13a}
\end{equation}

\begin{equation}
\frac{b_{+}}{a_{+}}=-\,\frac{ike^{-2ikL}\,W}{\left( u_{L}^{\prime
}-iku_{L}\right) \left( v_{L}^{\prime }-ikv_{L}\right) }  \label{13b}
\end{equation}

\noindent Agora focalizamos nossa aten\c{c}\~{a}o na determina\c{c}\~{a}o
dos coeficientes de reflex\~{a}o $R$ e transmiss\~{a}o $T$. O coeficiente de
reflex\~{a}o (transmiss\~{a}o) \'{e} definido como a raz\~{a}o entre as
correntes refletida (transmitida) e incidente. Haja vista que $\partial \rho
/\partial t=0$ para estados estacion\'{a}rios, temos que a corrente \'{e}
independente de $x$. Usando este fato obtemos prontamente que

\begin{eqnarray}
R &=&\frac{J_{\mathtt{ref}}}{J_{\mathtt{inc}}}=\frac{|a_{-}|^{2}}{|a_{+}|^{2}%
}  \nonumber \\
&&  \nonumber \\
&=&\frac{\left( k^{2}u_{L}\,v_{L}+u_{L}^{\prime }v\,_{L}^{\prime }\right)
^{2}}{\left( k^{2}u_{L}\,v_{L}-u_{L}^{\prime }\,v_{L}^{\prime }\right)
^{2}+k^{2}\left( u_{L}^{\prime }\,v_{L}+u_{L}\,v_{L}^{\prime }\right) ^{2}}
\label{15a}
\end{eqnarray}

\begin{eqnarray}
T &=&\frac{J_{\mathtt{trans}}}{J_{\mathtt{inc}}}=\frac{|b_{+}|^{2}}{%
|a_{+}|^{2}}  \nonumber \\
&&  \nonumber \\
&=&\frac{k^{2}W^{2}}{\left( k^{2}u_{L}\,v_{L}-u_{L}^{\prime }\,v_{L}^{\prime
}\right) ^{2}+k^{2}\left( u_{L}^{\prime }\,v_{L}+u_{L}\,v_{L}^{\prime
}\right) ^{2}}  \label{16a}
\end{eqnarray}%
Da\'{\i} o leitor pode mostrar que $R+T=1$, como deve ser por causa da
conserva\c{c}\~{a}o da probabilidade.

O formalismo desenvolvido acima tamb\'{e}m permite a an\'{a}lise de estados
ligados. Note que (\ref{5}), (\ref{7c}) e (\ref{8}) descrevem estados de
es\-pa\-lha\-men\-to com $E>0$ e $k$ $\in
\mathbb{R}
$. Poss\'{\i}veis estados ligados tamb\'{e}m poderiam ser descritos por
essas autofun\c{c}\~{o}es com $k=i|k|$, onde $|k|=\allowbreak \sqrt{%
2m|E|/\hbar ^{2}}$ com $E<0$, e $a_{+}=b_{-}=0$. Devemos impor que $a_{+}$ e
$b_{-}$ sejam nulos para que a densidade de probabilidade seja finita em $%
x=\pm \infty $ . Ora, tem que ser assim, pois $\int_{-\infty }^{+\infty
}dx\,|\psi |^{2}<\infty $. Para $|x|<L$ pode-se deduzir que $J=0$, e
portanto dever\'{\i}amos concluir que $\textrm{Im}\left( c_{P}^{\ast
}\,c_{I}\right) =0$, para se p\^{o}r de acordo com a equa\c{c}\~{a}o da
continuidade e com a express\~{a}o do fluxo na regi\~{a}o $|x|<L$ expressa
por (\ref{jota}). Nesta circunst\^{a}ncia, as rela\c{c}\~{o}es (\ref{C1}) e (%
\ref{C2}) fornecem%
\begin{equation}
u_{L}^{\prime }+|k|u_{L}=0,\quad c_{I}=0,\quad b_{+}=a_{-}=c_{P}\,u_{L}\,e^{-%
\frac{u_{L}^{\prime }}{u_{L}}L}  \label{cc1}
\end{equation}%
e%
\begin{equation}
v_{L}^{\prime }+|k|v_{L}=0,\quad c_{P}=0,\quad
b_{+}=-a_{-}=c_{I}\,v_{L}\,e^{-\frac{v_{L}^{\prime }}{v_{L}}L}  \label{cc2}
\end{equation}%
As autofun\c{c}\~{o}es correspondentes a (\ref{cc1}) e (\ref{cc2}) podem ser
escritas como%
\begin{equation}
\psi \left( x\right) =b_{+}\left\{
\begin{array}{c}
+e^{+|k|x} \\
\\
\frac{e^{-|k|L}}{u_{L}}\,u\left( x\right)  \\
\\
+e^{-|k|x}%
\end{array}%
\begin{array}{c}
{\textrm{para }}x<-L \\
\\
{\textrm{para }}|x|<L \\
\\
{\textrm{para }}x>L%
\end{array}%
\right.
\end{equation}%
para $\psi $ par, e
\begin{equation}
\psi \left( x\right) =b_{+}\left\{
\begin{array}{c}
-e^{+|k|x} \\
\\
\frac{e^{-|k|L}}{v_{L}}\,v\left( x\right)  \\
\\
+e^{-|k|x}%
\end{array}%
\begin{array}{c}
{\textrm{para }}x<-L \\
\\
{\textrm{para }}|x|<L \\
\\
{\textrm{para }}x>L%
\end{array}%
\right.
\end{equation}%
para $\psi $ \'{\i}mpar. Fortuitamente, as condi\c{c}\~{o}es para a exist%
\^{e}ncia de estados ligados tamb\'{e}m poderiam ser obtidas por meio da
identifica\c{c}\~{a}o dos polos das amplitudes expressas por (\ref{13a}) e (%
\ref{13b}) se os valores f\'{\i}sicos do n\'{u}mero de onda $k$, definidos
no eixo real, forem estendidos para o plano complexo. Com efeito, a prescri%
\c{c}\~{a}o $k\rightarrow i|k|$ anula o denominador de (\ref{13a}) e (\ref%
{13b}) sempre que%
\begin{equation}
u_{L}^{\prime }+|k|u_{L}=0  \label{quant1}
\end{equation}%
ou%
\begin{equation}
v_{L}^{\prime }+|k|v_{L}=0  \label{quant2}
\end{equation}%
As equa\c{c}\~{o}es expressas por (\ref{quant1}) e (\ref{quant2}) s\~{a}o
rela\c{c}\~{o}es impl\'{\i}citas para a determina\c{c}\~{a}o das poss\'{\i}%
veis autoenergias. A primeira fornece autovalores associados com autofun\c{c}%
\~{o}es pares ($c_{I}=0$), e a segunda com autofun\c{c}\~{o}es \'{\i}mpares (%
$c_{P}=0$). Da\'{\i} se v\^{e} que o fluxo de pro\-ba\-bi\-li\-da\-de
expresso por (\ref{jota}) \'{e} nulo no caso de estados ligados, como deve
ser.

Formalmente, tanto o problema de espalhamento quanto o problema de estados
ligados est\~{a}o resolvidos. Na se\c{c}\~{a}o seguinte ilustramos a t\'{e}%
cnica com o caso simples de um potencial retangular constitu\'{\i}do de tr%
\^{e}s patamares que reduz-se a ao po\c{c}o de potencial ou \`{a} barreira
de potencial, conforme o sinal de $\mathcal{V}(x)$.

\section{O potencial quadrado}

Consideremos agora
\begin{equation}
\mathcal{V}(x)=V_{0}  \label{quad}
\end{equation}%
e o n\'{u}mero de onda $q$ definido por%
\begin{equation}
q=\sqrt{\frac{2m}{\hbar ^{2}}\left( E-V_{0}\right) }
\end{equation}%
A segrega\c{c}\~{a}o dos casos $E>V_{0}$ e $E<V_{0}$, correspondendo a $q$
real e $q$ imagin\'{a}rio respectivamente, conduz a duas classes distintas
de solu\c{c}\~{o}es. A seguir, calcularemos explicitamente o coeficiente de
transmiss\~{a}o e encontraremos as condi\c{c}\~{o}es de quantiza\c{c}\~{a}o
para cada uma dessas classes de solu\c{c}\~{o}es, seja $V_{0}$ positivo ou
negativo.

\begin{itemize}
\item \underline{$E>V_{0}$}. Neste caso $q$ \'{e} um n\'{u}mero real e a equa%
\c{c}\~{a}o de Schr\"{o}dinger independente do tempo admite as solu\c{c}\~{o}%
es linearmente independentes%
\begin{equation}
u=\cos (qx)\quad \textrm{e}\quad v=\mathrm{sen}(qx)  \label{sol1}
\end{equation}%
Desta maneira, o wronskiano das solu\c{c}\~{o}es $u$ e $v$ \'{e} igual a $q$
e (\ref{16a}) torna-se%
\begin{equation}
T=\left\{ 1+\left[ \frac{k^{2}-q^{2}}{2kq}\,\mathrm{sen}(2qL)\right]
^{2}\right\} ^{-1}  \label{t1}
\end{equation}%
Ao passo que as condi\c{c}\~{o}es de quantiza\c{c}\~{a}o, expressas por (\ref%
{quant1}) e (\ref{quant2}), tornam-se
\begin{equation}
\frac{|k|}{q}=\left\{
\begin{array}{c}
\tan \left( qL\right) \\
\\
-\cot \left( qL\right)%
\end{array}%
\begin{array}{c}
{\textrm{para }}c_{I}=0 \\
\\
{\textrm{para }}c_{P}=0%
\end{array}%
\right.  \label{p1}
\end{equation}%
Conv\'{e}m lembrar que o coeficiente de transmiss\~{a}o s\'{o} \'{e} v\'{a}%
lido para $E>0$ ($k$ \'{e} real). Entretanto, as condi\c{c}\~{o}es de
quantiza\c{c}\~{a}o s\~{a}o v\'{a}lidas somente para $E<0$ ($k$ \'{e} imagin%
\'{a}rio puro), o que imp\~{o}e naturalmente que $V_{0}$ seja negativo. Em
outras palavras, somente o po\c{c}o de potential tolera a exist\^{e}ncia\ de
estados ligados.

\item \underline{$E<V_{0}$}. Neste caso $q$ \'{e} um n\'{u}mero imagin\'{a}%
rio puro e as solu\c{c}\~{o}es linearmente independentes s\~{a}o
\begin{equation}
u=\cosh (|q|x)\quad \textrm{e}\quad v=\mathrm{senh}(|q|x)
\end{equation}%
Desta feita, $W=|q|$ e
\begin{equation}
T=\left\{ 1+\left[ \frac{k^{2}+|q|^{2}}{2k|q|}\,\mathrm{senh}(2|q|L)\right]
^{2}\right\} ^{-1}  \label{t2}
\end{equation}%
Neste caso de energias menores que a altura do potencial, necessariamente
com $V_{0}>0$ e $E>0$, revela-se o efeito t\'{u}nel. Uma circunst\^{a}ncia
em que, embora n\~{a}o haja ondas progressivas na regi\~{a}o do potencial, h%
\'{a} uma corrente dada por%
\begin{equation}
J=\frac{\hbar |q|}{m}\,\textrm{Im}\left( c_{P}^{\ast }\,c_{I}\right)
\label{9e}
\end{equation}%
que se anula somente quando $k$ \'{e} um n\'{u}mero imagin\'{a}rio. As condi%
\c{c}\~{o}es de quantiza\c{c}\~{a}o (\ref{quant1}) e (\ref{quant2}) ditam que%
\begin{equation}
-\,\frac{|k|}{|q|}=\left\{
\begin{array}{c}
\tanh \left( |q|L\right) \\
\\
\coth \left( |q|L\right)%
\end{array}%
\begin{array}{c}
{\textrm{para }}c_{I}=0 \\
\\
{\textrm{para }}c_{P}=0%
\end{array}%
\right.  \label{95}
\end{equation}%
Contudo, estas condi\c{c}\~{o}es n\~{a}o fornecem solu\c{c}\~{o}es porque o
membro esquerdo de (\ref{95}) \'{e} negativo e os membros direitos s\~{a}o
positivos. Em outras palavras, a exist\^{e}ncia de estados ligados requer um
n\'{u}mero de onda real na regi\~{a}o do potencial. A aus\^{e}ncia de
estados ligados, verificada aqui em decorr\^{e}ncia das condi\c{c}\~{o}es de
quantiza\c{c}\~{a}o expressas por (\ref{95}), se d\'{a} porque as solu\c{c}%
\~{o}es normaliz\'{a}veis da equa\c{c}\~{a}o de Schr\"{o}dinger requerem que
$E$ exceda o m\'{\i}nimo de $V\!\!\left( x\right) $.
\end{itemize}

\bigskip

\section{Conclus\~{a}o}

Apresentou-se um formalismo que pode descrever estados de espalhamento e
estados ligados de uma forma unificada. O m\'{e}todo foi desenvolvido para
potenciais localizados sim\'{e}tricos mas pode ser estendido para potenciais
assim\'{e}tricos com relativa facilidade. O exemplo do afamado potencial
quadrado poderia nos conduzir a concluir que o m\'{e}todo \'{e} extremamente
poderoso, mas n\~{a}o \'{e} bem assim. Acontece que certas formas de $%
\mathcal{V}(x)$, ainda que sejam simples, deixam a proposta na berlinda
devido \`{a} equa\c{c}\~{a}o de Schr\"{o}dinger n\~{a}o resultar em solu\c{c}%
\~{o}es amig\'{a}veis para $u(x)$ e $v(x)$, e at\'{e} mesmo n\~{a}o redundar
em solu\c{c}\~{o}es anal\'{\i}ticas. Formas simples para $\mathcal{V}(x)$
com interesse pr\'{a}tico, por exemplo, incluem o potencial parab\'{o}lico
\cite{cru} e o potencial triangular \cite{ami}. As solu\c{c}\~{o}es da equa%
\c{c}\~{a}o de Schr\"{o}dinger para a primeira forma envolvem fun\c{c}\~{o}%
es hipergeom\'{e}tricas confluentes enquanto a segunda forma envolvem fun%
\c{c}\~{o}es de Airy. N\~{a}o obstante, o m\'{e}todo pode se tornar um
excelente ponto de partida para a busca de solu\c{c}\~{o}es num\'{e}ricas,
ou ainda de solu\c{c}\~{o}es anal\'{\i}ticas aproximadas, para o coeficiente
de transmiss\~{a}o e para as energias dos poss\'{\i}veis estados ligados.

\bigskip

\bigskip

\bigskip

\bigskip

\noindent \textbf{Agradecimentos:}

\medskip

\noindent O autor \'{e} grato ao CNPq pelo apoio financeiro.

\end{document}